\documentclass[12pt]{article}
\usepackage{graphicx}
\begin{document}

\hsize 14.5 cm
\vsize 21 cm

\leftline {\LARGE Quantum Spin Glass Phase Boundary in $\pm J$}

\medskip

\leftline {\LARGE   
Transverse Field Ising Systems\footnote{
Presented at CMDays-03, Jadavpur Univ., Kolkata, Aug. 2003. Proc. in Ind. J. 
Phys. (to be published).}}

\vspace{0.6cm}

\noindent Arnab Das$^\dagger$, Amit Dutta$^\ddagger$ and 
Bikas K. Chakrabarti$^\dagger$

\vspace{0.6cm}

\noindent $^\dagger$ Saha Institute of Nuclear Physics,
1/AF Bidhannagar, Kolkata - 700064, India \\

\noindent $^ \ddagger$ Physics Department, Indian Institute of
Technology, Kanpur - 208016, India

\vspace{1.3cm}
 
\noindent{\bf Abstract:}\quad Here we study zero temperature 
quantum phase transition driven by the transverse field for random $\pm J$
Ising model on chain and square lattice.  
We present some analytical results for one dimension and some numerical results
for very small square lattice under periodic boundary condition.
The numerical results are obtained employing exact diagonalization technique 
following Lanczos method. \\

\vspace{0.6cm}


\vspace{0.6cm}


\vspace{0.6cm} 

\noindent {\bf 1.\qquad Introduction} 

\medskip
\noindent  
The interest in the
study of  transverse Ising spin glass models was revived in early
 1990 by the discovery of zero-temperature transition in dipolar Ising 
transverse field magnet LiHo$_x$Y$_{1-x}$F$_4$ [1]. 
Proton glasses such as mixture of ferroelectric and
anti-ferroelectric materials like  Rb$_{(1-x)}$(NH$_4$)$_x$(H$_2$P)$_4 $
 [2] also provided earlier useful realizations of such
 quantum spin 
glasses. \\

\noindent These developments initiated extensive theoretical 
studies in quantum spin glass models.
Ising model in transverse field has already been studied 
extensively in this context
through analytical approaches using approximate renormalization 
techniques and real space
renormalization group method, as well as using numerical methods 
like quantum Monte Carlo
 and exact diagonalization techniques [3]. Fairly extensive studies 
on the quantum spin glass
phases have been made so far for transverse field  Edward-Anderson
 model and 
Sherrington-Kirkpatrick model with
random exchange distributions [3, 4] and to some extent on quantum 
Heisenberg spin glass models
[4]. In view of the rigorous developments in the study of two 
dimensional nearest neighbour Ising model with random $\pm J$ 
exchange interactions  and the precise knowledge of location 
of the Nishimori line [5] in
such classical spin glass model (driven by temperature), we 
consider here the quantum phase
transition (at zero temperature) in the same $\pm J$ Ising model.\\

\noindent We have shown analytically that introduction of random $-J$ 
impurities cannot affect the zero temperature phase transition in  
one dimensional system as they can be transformed away.
We have also compared and verified the result numerically
for the small system size considered.
For two dimensional systems,
  we present some priliminary results obtained for a square lattice using
exact diagonalization results for very small system sizes following Lanczos 
technique [6]. Only the behaviours of configurationally averaged
 energy gap $\Delta=(E_1 - E_0)$  between the first excited state   
and  the ground state  
 and the second order response function 
$\chi = \left( \partial^2 E_0  / \partial\Gamma^2 \right)$, 
equivalent to specific
heat, have been studied here. The variations of $\Delta$ and $\chi$ 
  with 
respect to transverse field $\Gamma$
have been obtained, and the phase 
boudary has been estimated from these results. \\

\noindent We work with a transverse Ising
system, using only nearest neighbour interactions, whose Hamiltonian  
is given by
\begin{equation}
{\mathcal H} = -\sum_{\langle i,j \rangle } J_{ij} S_i^z S_j^z - 
\Gamma\sum_{i=1}^N S_i^x ,
\end{equation} 
\noindent where the transverse field $\Gamma$ is uniform through out the system
and the nearest neighbour exchange constant $J_{ij}$'s are chosen randomly from the binary distribution
\begin{equation}
P(J_{ij}) = p\delta(J_{ij} + J) + (1 - p)\delta(J_{ij} - J).
\end{equation}
\noindent Here $J$ is taken positive and $p$ is thus the concentration of 
anti-ferromagnetic $-J$ bonds in the system.\\

\noindent {\bf 2. \qquad Results in One Dimensional System }

\medskip 
\noindent Here first we show analytically
  that in a one dimensional transverse Ising 
Hamiltonian with uniform $J$ and $\Gamma$,
if we replace some $J$ bonds by $-J$ bonds randomly, then the resulting 
Hamiltonian can be gauge transformed back to one with uniform $J$, 
and hence the critical field remains unchanged with randomness concentration. 
Simillar result for one dimensional system with distributed $J$ 
had been obtained earlier [7]. \\

\noindent Let us take the one dimensional random bond tranverse Ising 
Hamiltonian 
\begin{equation} 
{\mathcal H} =   
-\sum_i J_i S_i^z S_{i+1}^z - \sum \Gamma S_i^x, 
\end{equation} 
where the transverse field $\Gamma$ is uniform throughout the system, and $J_{i}$'s are randomly
chosen from the same distribution as given in $(2)$. 
 Since the $J_i$'s have same magnitude $J$ all through, 
and their randomness is
only in their sign, we may write $ J_i = J {\mathrm {sgn}}(J_i)$ , 
and thus Hamiltonian (3)
takes the form 
\begin{equation}
 {\mathcal H} = 
-J\sum {\mathrm {sgn}}(J_i) S_i^z S_{i+1}^z - \Gamma\sum  S_i^x . 
\end{equation}

\noindent Now let us  define a new set of spin variables as below

$$ \tilde{S}_i^z = S_i^z \prod_{k=1}^{i-1} {\mathrm {sgn}}(J_k) $$    

$$ \tilde{S}_i^x = S_i^x $$

$$ \tilde{S}_i^y = S_i^y \prod_{k=1}^{i-1} {\mathrm {sgn}}(J_k). $$
 It is easy to see that $\tilde{S}$ 's satisfy the same commutation and 
anti-commutation relations as those of $S$'s and hence will exhibit exactly
the same dynamical behaviour. Now,

$$ \tilde{S}_i^z\tilde{S}_{i+1}^z 
= S_i^z S_{i+1}^z \left[\prod_{k=1}^{i-1}[{\mathrm {sgn}}
(J_k)]^2 \right]{\mathrm {sgn}}(J_i), $$
      
\noindent or,

$$ \tilde{S}_i^z\tilde{S}_{i+1}^z = S_i^z S_{i+1}^z {\mathrm  {sgn}}(J_i), $$ 
since $ [{\mathrm {sgn}}(J_k)]^2 = 1 $. 
Thus in terms of new spin variables, Hamiltonian 
(4) becomes
\begin{equation}
{\mathcal H} = -J\sum_i \tilde{S}_i^z \tilde{S}_{i+1}^z - 
\Gamma \sum_i \tilde{S}_i^x. 
\end{equation}
              The above Hamiltonian describes the same random 
system in terms of
new variables, and yet, as one can see, it has in itself no randomness at all. 
One can use Jordon-Wigner transformation in terms of $\tilde{S}$'s and see that
here also
quantum phase transition occurs only at $\Gamma \ge \Gamma_c (=J)$ as it occurs
in a non-random Hamiltonian in $S$'s. In Fig. 1, we present 
some data computed for a
chain of size $N=9$, which shows that the gap $\Delta$ vanishes 
 at $\Gamma_c  \approx 1$ (the field being scaled by $J$). These data for
$\Delta = E_1 - E_0$ is obtained from the computed average value of $E_0$ and
 $E_1$, each one averaged over about 10 configurations for $p \ne 0$.
For infinite system, $\Delta$ is a linear function of $\Gamma$ for 
$\Gamma \ge \Gamma_c$. In our case, linearity is observed at high values of
$\Gamma$, and $\Gamma_c$ is determined by backward linear extrapolation from
the linear region.
\begin{figure}[htb]
\resizebox{13.0cm}{!}{\rotatebox{270}{\includegraphics{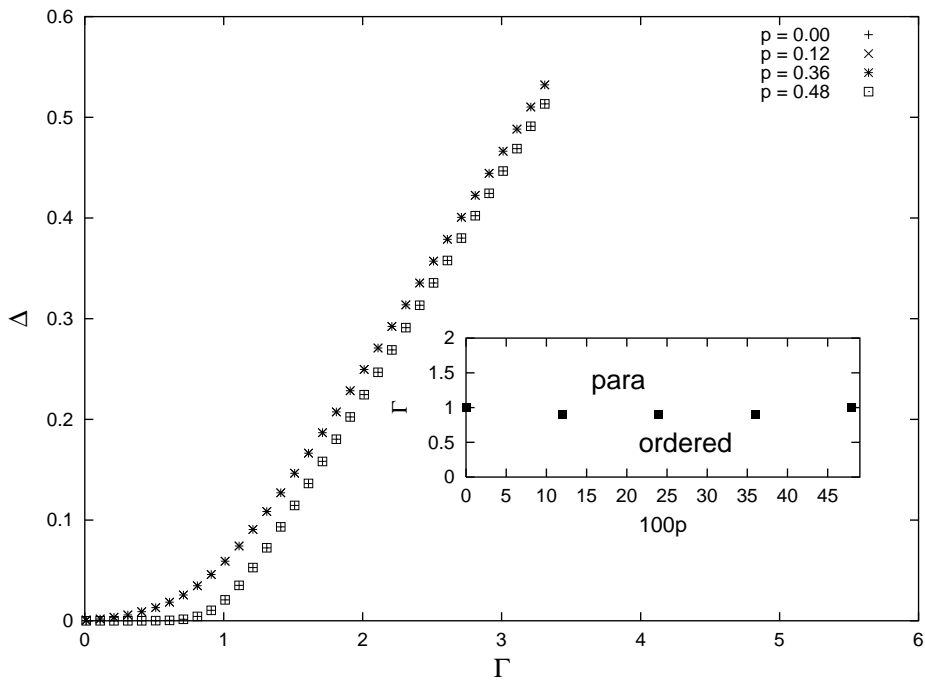}}}  
\caption{A numerical estimate of the energy gap $\Delta$ 
for a chain with $ N $= 9. Phase boundary is obtained from the 
location of $\Delta(\Gamma) = 0$, and is shown in the inset.}
\end{figure}
 

\noindent From the numerical data in Fig. 1, we see that there is a slight
variation of $\Gamma_c$ with $p$ ($\Gamma_c$ varies between 0.9 and 1.0). This
variation can be attributed to the very small size of the system. However, 
it may be noted that with even number of $-J$ bonds in the 
chain, with periodic boundary condition, there is no problem of 
incommensuration and $E_0 (\Gamma)$ or $E_1 (\Gamma)$ become strictly identical
for such values of $p$. Similarly, in every case of odd number of 
$-J$ bonds in the 
chain, incommensuration problem always occurs for one spin only, rendering
identical values (but different from the even $-J$ case) 
for $E_0 (\Gamma)$ and $E_1 (\Gamma)$
in all such cases. \\

\noindent {\bf{3. \qquad  Results for Square Lattice}}

\medskip

\noindent We consider now the same system (represented by Hamiltonian (1)) on
a square lattice of size 3$\times$3 with periodic boundary condition. We again 
calculate the ground state and the first excited state energy $E_0$ and $E_1$ 
respectievely as functions of the transverse field $\Gamma$, for different
values of $p$. Each value of $E_0$ and $E_1$ is averaged over at least 10 
configurations for each $p \ne 0$. Apart from $\Delta$, we also calculate
$\chi = \partial^2E_0/\partial\Gamma^2$ and their variations with $\Gamma$
as shown in Figs. 2 and 3 respectively.
\begin{figure}[htb]
\resizebox{13.0cm}{!}{\rotatebox{270}{\includegraphics{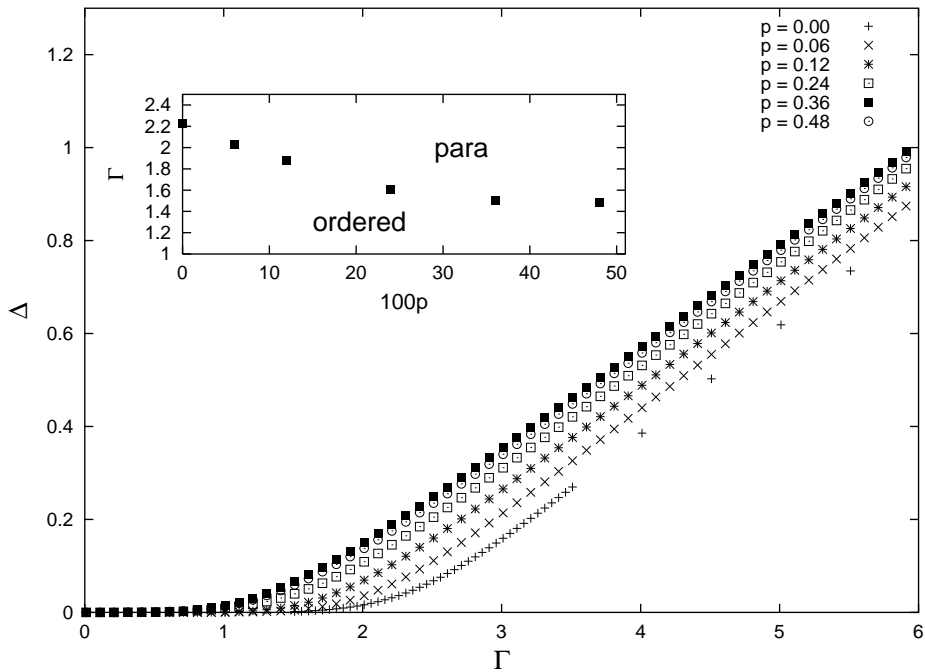}}}
\caption{A numerical estimate of the configurational avrage of the energy
gap $\Delta$
for a square lattice of size 3$\times$3. Phase boundary
obtained from $\Delta (\Gamma) = 0$ is outlined
 in the inset.}
\end{figure}

\noindent Our results here are severely 
constrained by the system size. The value of pure ferromagnetic critical field 
$\Gamma_c(p=0)$ is found here to be around 2.2, while the series 
results [3] or cluster algorithms [8] give the value to be around 3.0. This 
discrepancy is attributed to the smallness of our system size 
$(N = 3^2 )$. However the qualitative behaviour of the order-disorder phase
boundary (between ferro/spin glass and para) seems to be reasonable: 
$\Gamma_c(p)$ decreases with $p$ initially, and then increases again as $p$
approaches unity (pure anti-ferromagnet). The use of periodic boundary 
condition here (to avoid some numerical errors) also restricts the domain 
features and thereby affects our results. The absence of the knowledge of
the ground-state wave function (and the correlation functions) in this method
also forbids us to analyse the structure of the ordered phases. \\

\begin{figure}
\resizebox{12.0cm}{!}{\rotatebox{270}{\includegraphics{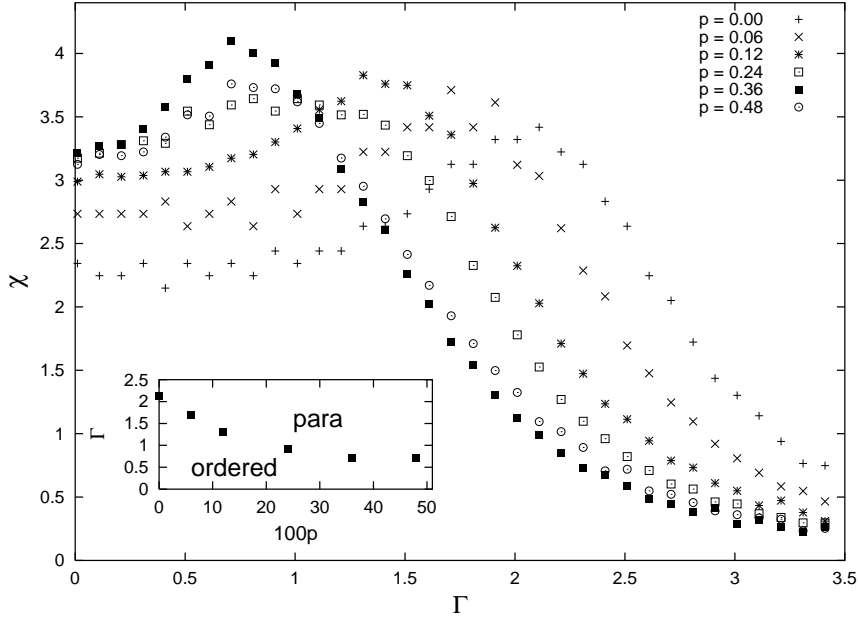}}}
\caption{Here variation of $\chi = \partial^2 E_0 / \partial \Gamma^2 $ with
$\Gamma$ is shown. The transition point occurs at $\Gamma = \Gamma_c$ where 
$\chi$ diverges; for finite system one gets only a  peak in $\chi$ at 
$\Gamma = \Gamma_c(p)$.
We have outlined the corrosponding phase boundary in the inset. }
\end{figure}
\vskip 0.6 cm

\noindent {\bf Acknowledgement:} BKC is grateful to Hidetoshi Nishimori 
for useful discussions. \\\
\vskip 1 cm

\noindent {\LARGE\textbf{References}} \\
\vspace{0.7cm}

\noindent [1] W. Wu, B. Ellmann, T.F. Rosenbaum, G. Appeli and D.H. 
Reich, Phys. 
Rev. Lett. {\bf 67} 2076 (1991);  
 W. Wu, D. Bitko, T.F. Rosenbaum and G. Appeli, Phys. Rev. Lett. 
{\bf 71} 1919 (1993) \\

\noindent [2] R. Pric, B. Tadic and R. Blinc, Z. Phys. B {\bf 61} 69 (1985) \\

\noindent [3] B. K. Chakrabarti, A. Dutta and P. Sen,         
{\it Quantum Ising Phases and Transitions in Transverse Ising Models}, 
Lecture Notes in Physics, {\bf m41}, Springer, Heidelberg (1996)\\

\noindent [4] R. N. Bhatt, in {\it Spin Glasses and Random Fields}, Ed.
A. P. Young, p. 225, World Sc., Singapore (1998) \\

\noindent [5] H. Nishimori, Prog. Theor. Phys. {\bf 66} 1169 (1980);
 {\it Statistical Physics of Spin Glasses \& Information Processing : 
An Introduction}, Oxford University Press, Oxford (2001); 
H. Nishimori and K. Nemoto, J. Phys. Soc. Jpn. {\bf 71} 1198 (2002) \\

\noindent [6] J. Stoer, R. Bulirsch, {\it Introduction to Numerical Analysis},
Text in Appl. Maths. {\bf12}, Springer-Verlag, New York, (1993 ) \\

\noindent [7] B. McCoy, in {\it Phase Transitions and Critcal Phenomena}, vol.
II, Eds. C. Domb and M. S. Green, Academic Press, London (1983);   
D. S. Fisher, Phys. Rev. B {\bf 50} 3799 (1994) \\

\noindent [8] H. Rieger and N. Kawashima, Euro. Phys. J. B {\bf 9} 233 (1999)

\end{document}